\documentclass[aps,prl,twocolumn,showpacs,amsmath,amssymb,floatfix,superscriptaddress]{revtex4}
\usepackage{graphicx}
\usepackage{bm}
\usepackage{epsfig}

\def\be{\begin{equation}}
\def\ee{\end{equation}}
\def\bea{\begin{eqnarray}}
\def\eea{\end{eqnarray}}
\def\la{\langle}
\def\ra{\rangle}

\begin{document}
\title{Quantum Ghost Imaging through Turbulence}

\author{P.\ Ben Dixon}
\affiliation{Department of Physics and Astronomy, University of Rochester, Rochester, New York 14627, USA}

\author{Gregory Howland}
\affiliation{Department of Physics and Astronomy, University of Rochester, Rochester, New York 14627, USA}

\author{Kam Wai Clifford Chan}
\affiliation{Rochester Optical Manufacturing Company, 1260 Lyell Avenue, Rochester, NY 14606, USA}

\author{Colin O'Sullivan-Hale}
\affiliation{Institute of Optics, University of Rochester, Rochester, New York, 14627, USA}

\author{Brandon Rodenburg}
\affiliation{Institute of Optics, University of Rochester, Rochester, New York, 14627, USA}

\author{Nicholas D. Hardy}
\affiliation{Research Laboratory of Electronics, Massachusetts Institute of Technology, Cambridge, MA 02139, USA}

\author{Jeffrey H.\ Shapiro}
\affiliation{Research Laboratory of Electronics, Massachusetts Institute of Technology, Cambridge, MA 02139, USA}

\author{D.\ S.\ Simon}
\affiliation{Department of Electrical \& Computer Engineering, Boston University, Boston, MA 02215, USA}

\author{A.\ V.\ Sergienko}
\affiliation{Department of Electrical \& Computer Engineering, Boston University, Boston, MA 02215, USA}

\author{R.\ W.\ Boyd}
\affiliation{Department of Physics and Astronomy, University of Rochester, Rochester, New York 14627, USA}
\affiliation{Institute of Optics, University of Rochester, Rochester, New York, 14627, USA}

\author{John C.\ Howell}
\affiliation{Department of Physics and Astronomy, University of Rochester, Rochester, New York 14627, USA}

\date{\today}

\begin{abstract}
We investigate the effect of turbulence on quantum ghost imaging.  We use entangled photons and demonstrate that for a novel experimental configuration the effect of turbulence can be greatly diminished.  By decoupling the entangled photon source from the ghost imaging central image plane, we are able to dramatically increase the ghost image quality.  When imaging a test pattern through turbulence, this method increased the imaged pattern visibility from \( \mathcal{V} = 0.14 \pm 0.04\) to  \( \mathcal{V} = 0.29 \pm 0.04\).
\end{abstract}


\maketitle

{\it Introduction.}--- The phenomenon of ghost imaging (GI), first observed by Pittman et al.\ in 1995 \cite{PittmanGI}, is a method of generating the image of an object from correlation measurements. Pittman's experiment made use of pairs of entangled photons.  One of the photons passed through a transmission object and then to a photon counter with no spatial resolution.  The other photon passed directly to a spatially resolving photon counter.  When looking at coincident photon detections, the detectors were able to see the object despite the fact that the object and the spatially resolving detector were in different arms of the experiment.  While it was initially thought to be a quantum mechanical effect reliant upon the entanglement between the two photons, similar results were later obtained using classical sources \cite{ThermalGI}.

In addition to clarifying the boundary between quantum and classical effects \cite{Q-C-GI1, Q-C-GI2, Q-C-GI3}, GI has been used for lensless imaging \cite{LenslessImaging}, super-resolution imaging \cite{SuperResolution1, SuperResolution2}, and entanglement detection \cite{EntDet}.  More recently, research has recognized connections between GI and compressive sensing \cite{CompGI, CSGI}.  The distributed nature of GI has made it a candidate for distributed image processing, and for distributed sensing and communication schemes.

For many optical applications, imaging through turbulence is unavoidable \cite{TurbBook, TurbFante}. GI is no different and the effect of turbulence on GI performance has begun to be theoretically investigated \cite{Turb1, Turb2, Turb3}.  In this paper, we experimentally investigate the effect of turbulence on GI using entangled photons.   We introduce a novel GI setup that allows us to minimize the effect of turbulence.

\begin{figure}
\includegraphics[scale=0.32]{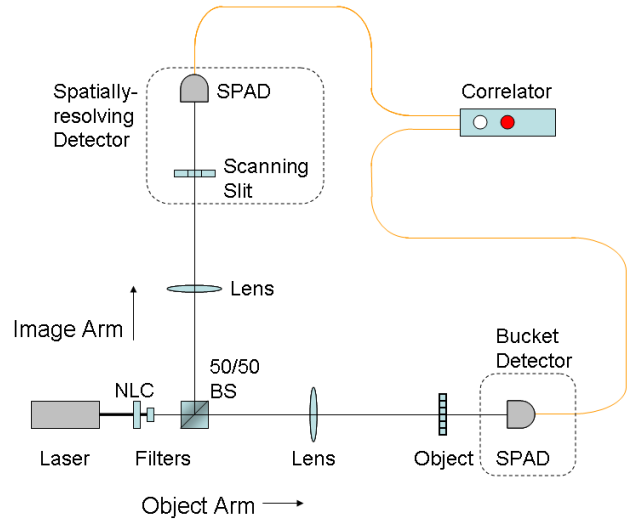}
\caption{(color online).  A \(325\) nm pump beam undergoes SPDC at a nonlinear crystal (NLC), the output passes a beamsplitter (BS). One beam is sent through a lens and onto a transmission object.  The other beam is sent through a lens and onto a scanning slit.  The ghost image of the object is profiled by the slit. Photons are detected with single-photon avalanche diodes (SPAD). A coincidence circuit correlates the measurements. }
\label{Experiment}
\end{figure}

\begin{figure}
\includegraphics[scale=0.27]{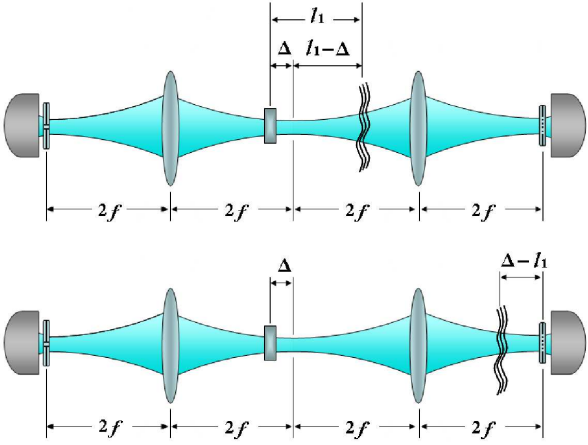}
\caption{(color online).  The experimental configuration is shown conceptually using the Klyshko picture \cite{klyshko82}, the object mask (on the right) is ghost imaged onto the scanning slit (on the left).  The nonlinear crystal is offset from the central image plane by a distance \(\Delta\).  The top picture shows the turbulence---represented by wavy lines---between the crystal and the lens.  The bottom picture shows the turbulence located between the lens and the object.  Experimentally relevant distances are labelled.}
\label{Klyshko}
\end{figure}

{\it Theoretical description.}---The experimental apparatus is depicted in Fig.\ \ref{Experiment}.  A biphoton state \( |\psi \ra \) is created at a nonlinear crystal \cite{klyshko88} and then split by a 50/50 beamsplitter, sending the biphoton into two arms of the apparatus.

In the object arm, the biphoton travels a distance \( 2f+\Delta \), to a lens which has focal length \(f\). The biphoton then travels a distance \(2f\) to a photon detector with no spatial resolution (a ``bucket'' detector).  A transmission object---consisting of alternating opaque and clear vertical bars---is placed just in front of the detector.

In the image arm, the biphoton travels a distance \( 2f-\Delta \) to a lens which again has focal length \(f\). The biphoton then travels a distance \(2f\) to a spatially-resolving photon detector.

For \(\Delta = 0 \) the detectors and crystal are all located at image planes of each other.  As one arm's lens/detector is moved {\it towards} the crystal by a distance \( \Delta \), the other arm's lens/detector is moved {\it away} by the same distance, keeping the sum of the arm's length constant, see Fig.\ \ref{Klyshko}.

Turbulent air flow is introduced into a thin region in the beam path of the object arm. For turbulence between the crystal and the lens, it is a distance \(l_1 \) from the crystal---or a distance \(l_1-\Delta \) from the central image plane.  For turbulence between the lens and the object, it is a distance \( \Delta - l_1 \) from the object.

The relevant function for GI is the second order degree of coherence  \(\mathrm{G}^{(2)}(x_1, x_2) \), where \(x_1\) is a transverse position variable in the plane of the spatially-resolving detector and \(x_2\) is a transverse position variable in the plane of the bucket detector.  We begin with the standard quantum mechanical form and include an additional ensemble averaging---represented by large outer brackets---to account for the statistical effect of turbulence:
\be
\mathrm{G}^{(2)}(x_{1},x_{2}) = \bigg\la \la \psi | \mathrm{{\hat E }}^{\dagger}_{i}(x_{1})\mathrm{{\hat E }}^{\dagger}_{s}(x_{2})\mathrm{{\hat E }}_{s}(x_{2})\mathrm{{\hat E }}_{i}(x_{1})  | \psi \ra \bigg\ra.
\label{doc1}
\ee
Neglecting overall normalization, this can be represented in the following way:
\be
\begin{split}
\mathrm{G}^{(2)}(x_{1},x_{2}) = \bigg\la \int^{(4)} \psi^{\star}(\tilde{x}_{s}, \tilde{x}_{i})
\mathrm{H}^{\star}(\tilde{x}_{i},x_{1})\mathrm{H}^{\star}(\tilde{x}_{s},x_{2}; \tilde{x}_t)\\
\times \mathrm{H}(x_{s},x_{2}; x_t)\mathrm{H}(x_{i},x_{1})
\psi(x_{s}, x_{i})
\mathrm{d}\tilde{x}_{i} \mathrm{d}\tilde{x}_{s}  \mathrm{d}x_{s} \mathrm{d}x_{i} \bigg\ra .
\label{doc2}
\end{split}
\ee
Subscript \(s\) and \(i\) indicate variables in the crystal plane and subscript \(t\) indicates variables in the plane of the turbulence.  The function \( \psi (x_s, x_i) \) is the transverse biphoton wavefunction which we approximate as a plane-wave with delta function correlations \( \psi (x_s,x_1) = \delta(x_s-x_i) \).  The function \( \mathrm{H}(x_s,x_2;x_t) \) is a propagation operator going from the crystal plane to the object arm detection plane, passing through the plane of turbulence; \(\mathrm{H}(x_i,x_1) \) is a propagation operator going from the crystal plane to the image arm detection plane.  These operators can be represented in the following way:
\be
\begin{split}
\mathrm{H}(x_s,x_2;x_t)= \int &\exp \left[ \frac{-i k\,(x_2-x_t)^2}{2 (l_1-\Delta)} \right] \mathrm{{\hat T}}(x_t) \\
 \times &\exp \left[ \frac{i k (x_t-x_s)^2\,}{2 l_1} \right] \mathrm{d}x_t,
\end{split}
\ee
\be
\mathrm{H}(x_i,x_1) = \exp \left[ \frac{-i k}{2 \Delta}(x_i-x_1)^2 \right].
\ee
We are assuming a narrow sheet of turbulent air, whose effect on propagation can be characterized by a multiplicative operator \( \mathrm{{\hat T}}(x_t) \).  In our theoretical treatment, we assume that the lenses are sufficiently large that they capture all of the light from the SPDC source. As a result, both turbulence locations in Fig.\ \ref{Klyshko} are governed by the same operators.

We model the turbulence as a \(6/3\) scaling law effect:  \( \big\langle  \mathrm{{\hat T}}^{\star}(\tilde{x}_t) \mathrm{{\hat T}}(x_t)  \big\ra =  \exp\left[ - \alpha \, (x_t-\tilde{x}_t)^2 /2 \right] \), where \( \alpha\) parameterizes the strength of the turbulence and has units \( 1/ \mathrm{m}^2 \) \cite{TurbBook, TurbFante}.  After integration, the resulting expression for \(\mathrm{G}^{(2)}(x_1,x_2) \) is:
\be
\mathrm{G}^{(2)}(x_{1},x_{2}) = \exp \left[ \frac{-  k^2 \big(x_1-x_2 \big)^2}{2 \alpha \left( l_1-\Delta \right)^2} \right].
\label{doc5}
\ee

The ghost image  \( \mathcal{I}(x_1) \) is then the product of the object and \( \mathrm{G}^{(2)}(x_{1},x_{2}) \), integrated over \( x_2 \).  We represent the object as: \(\mathcal{O}(x_2) = \exp \left[ - x_2^{\,2} /2 w^2 \right] \big( 1+ \cos(k_{o}\,x_2) \big) \).  Here \(w\) is the spatial width of the illuminating beam and \( k_{o} \) is wavenumber for the object's pattern spacing.

Assuming \( (l_1-\Delta) \sqrt{\alpha} \ll k\,w \), the ghost image is found to be
\be
\mathcal{I}(x_1) = \exp \left[-\frac{1}{2} \left(\frac{x_1}{w}\right)^2 \right] \Big( 1+\mathcal{V} \cos(k_{o}\,x_1) \Big).
\label{GIfinal}
\ee

\(  \mathcal{I}(x_1)  \) has the same form as \( \mathcal{O}(x_1) \) with the object's unity visibility replaced by the detected ghost image visibility \( \mathcal{V}\):
\be
\mathcal{V} = g \times \exp \left[ \frac{-\alpha\,\big(l_1-\Delta\big)^2}{2 \left( k / k_0 \right)^2} \right].
\label{V}
\ee
Where \(g\) is the the optimum ghost image visibility for turbulence.  As the turbulence increases in strength (increasing \(\alpha\)), the detected visibility \(\mathcal{V}\) decreases---thus obscuring the detected pattern.  Similarly, when the turbulence is moved away from either the central image plane or the detector, \( l_1-\Delta \) increases, thereby decreasing \(\mathcal{V}\).

{\it Experiment.}---Collimated light from a 3 mW, 325 nm HeCd laser with a 1/e\(^2\) full width of approximately 1600 \(\mu\)m  pumped a 10 mm thick BBO nonlinear optical crystal.  The crystal was oriented for degenerate type-I collinear spontaneous parametric down-conversion (SPDC).  After the crystal, the pump beam was blocked by colored glass filters and the SPDC bandwidth was limited by a 3nm wide spectral filter centered at 650 nm.  The remaining SPDC beam was split into two arms by a 50/50 beamsplitter.

In the image arm, a lens was located \(1000\,\mathrm{mm}-\Delta\) from the crystal; in the object arm, a lens was located \(1000\,\mathrm{mm}+\Delta\) from the crystal. Both lenses had focal length \(f = 500 \) mm.   Detectors were located 1000 mm from the lenses.

The transmission object was a test pattern located 1000 mm from the lens.  The bucket detector consisted of a \(10\times\) microscope objective which collected the transmitted light into a multimode optical fiber.  The pattern had unity visibility and 3.6 cycles per mm, which resulted in an object pattern wavenumber of \( k_o = 7.2\times\pi\) mm\(^{-2}\)

The spatially-resolving detector consisted of a computer controlled scanning slit located 1000 mm from the lens, which was again followed by a \(10\times\) microscope objective which collected the transmitted light into a multimode optical fiber.  The slit had an approximate width of 40 \(\mu\)m and was scanned in 5 \(\mu\)m increments, giving spatial resolution.

The optical fibers were connected to Perkin Elmer single-photon avalanche diode detectors.  The outputs of these detectors were time correlated using a PicoHarp 300 from PicoQuant.  Photon counts were integrated at each slit location for between 1 and 4 seconds.  The spatially resolved coincident detections made up the ghost image profiles.

A downward pointing heat gun was mounted above the setup, providing turbulent air flow across the beam path.  The effect of the turbulence was fitted to the model's wave structure function \( \alpha\,x^2 \) \cite{TurbFante}.  From the fit we determined \( \alpha = 2.5 \pm 1.5\,\mathrm{mm}^{-2} \).

Data was taken for an unshifted configuration with \(\Delta = 0 \), and for a shifted configuration with \(\Delta = 330\) mm.  In each configuration, ghost images were recorded with turbulence present in the object arm: both between the crystal and lens, and between the lens and the object. Ghost images were also recorded with no turbulence.

While allowing access to the central image plane of the apparatus, the shifted configuration introduced two experimental limitations: the flux through the object and scanning slit decreased significantly as a result of the detectors being away from the beam focus, and fewer spatial frequencies contributed to the ghost image as a result of the nonlinear crystal having a stronger aperturing effect.

The recorded ghost image profiles were fitted to \(\mathcal{I}(x_1) \) from Eq.\ \ref{GIfinal}.  The fit included a visibility term which constituted our measurement of the visibility \(\mathcal{V}\).

\begin{figure}
\includegraphics{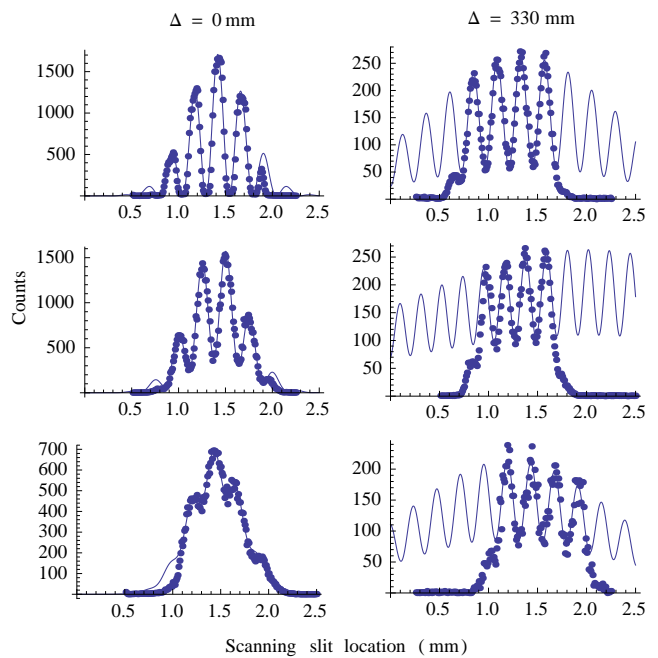}
\caption{(color online). Representative ghost images for the unshifted configuration (left), and shifted configuration (right).  The top row shows images with no turbulence.  The middle row shows images for turbulence between the lens and the object, 203 mm (right) and 229 mm (left) from the object.  The bottom row shows images for turbulence between the crystal and the lens, 432 mm from the crystal.  Points are experimental data while cuves are fits to the data.  Counts are measured in coincident photon detections per second.}
\label{RepScans}
\end{figure}

Representative ghost images are shown in Fig.\ \ref{RepScans}.  With no turbulence, the unshifted configuration produced ghost images, with visibilities of \(1.00 \pm 0.05 \).  With no turbulence, the shifted configuration produced ghost image visibilities of only \(0.65 \pm 0.05 \).  The scans also show the decreased flux and the broader beam profile associated with the shifted configuration.

\begin{figure}
\includegraphics{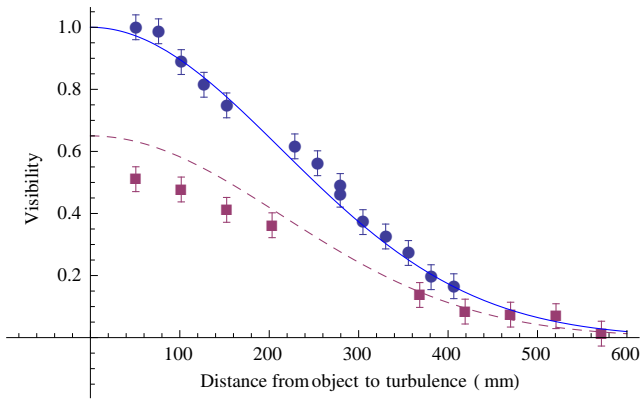}
\caption{(color online). Ghost image visibilities are shown for turbulence between the lens and the object.  Visibilities are plotted as a function of distance from the object to the turbulence (\(l_1 - \Delta\) in Eq.\ \ref{V}).  Data for the unshifted configuration are shown as blue circles while data for the shifted configuration are shown as purple squares.  Curves are plots from Eq.\ \ref{V}.  The solid curve is for the unshifted configuration, with \(g = 1.00\), and \(\Delta = 0\). The dashed curve is for the shifted configuration, with \(g = 0.65\) and  \(\Delta = 330\) mm.  For both curves \(\alpha = 2.0\, \mathrm{mm}^{-2} \).}
\label{Vis1}
\end{figure}

Visibilities for turbulence between the lens and the object are shown in Fig.\ \ref{Vis1}. For both the shifted and unshifted configurations, when the turbulence was close to the object, the observed visibility was near its no turbulence level.  As the turbulence was moved away from the object, the ghost image visibility decreased, thereby degrading the image.  The visibility for the unshifted configuration remained above the visibility for the shifted configuration for all turbulence locations.

\begin{figure}
\includegraphics{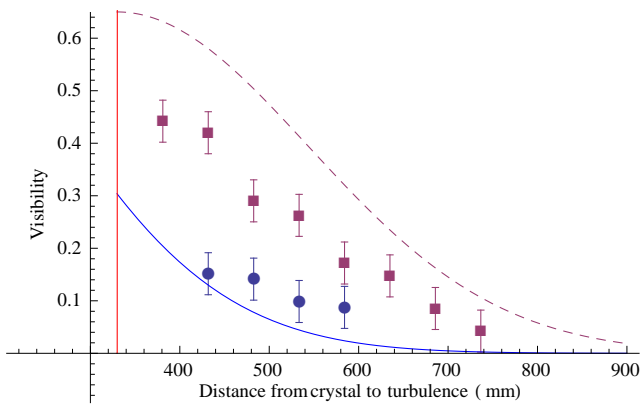}
\caption{(color online). Ghost image visibilities are shown for turbulence between the crystal and the lens.  Visibilities are plotted as a function of distance from the crystal to the turbulence (\(l_1\) in Eq.\ \ref{V}).  Data for the unshifted configuration are shown as blue circles while data for the shifted configuration are shown as purple squares.  Curves are plots from Eq.\ \ref{V}.  The solid curve is for the unshifted configuration, with \(g = 1.00\), and \(\Delta = 0\). The dashed curve is for the shifted configuration, with \(g = 0.65\) and  \(\Delta = 330\) mm.  For both curves \(\alpha = 2.0\, \mathrm{mm}^{-2} \).  The vertical line marks the location of the central image plane.}
\label{Vis2}
\end{figure}

Visibilities for turbulence between the crystal and the lens are shown in Fig.\ \ref{Vis2}. This is the main result of the experiment.  Visibilities decreased as the turbulence was moved away from the crystal, however, the unshifted configuration had {\it lower} fringe visibility than the shifted configuration.  Indeed, for turbulence located 482 mm from the crystal, the visibility was \( \mathcal{V} = 0.14 \pm 0.04\) for the unshifted configuration, while for the shifted configuration it was \( \mathcal{V} = 0.29 \pm 0.04\). Moving to the shifted configuration doubled the visibility.  Data was taken for turbulence as close as 380 mm to the crystal, or 50 mm from the central image plane.

{\it Concluding remarks.}---By moving the crystal from the central image plane we were able to place turbulence in this plane.  This decreased the observed effect of turbulence, in fact it more than made up for the inherent loss of visibility associated with the shifted configuration.  This technique has use in free space GI applications where turbulence is involved.  By arranging detectors to place an image plane at the location of the turbulence, image degradation from the turbulence can be diminished.

Although we have used entangled photons, similar results are expected for thermal light GI.  It should also be noted that the theoretical description assumes delta function correlations for the biphoton state and uses a thin region, non-Kolmogorov turbulence model \cite{TurbBook, TurbFante, KolmogorovTurb}.  We are currently extending our theoretical description to include different SPDC correlation areas and a more complex description of turbulence including the possibility of volume turbulence.  This will be presented in a forthcoming paper.

In this paper we have demonstrated a method of ameliorating the effects of turbulence on GI systems, and have provided a theoretical model which accurately describes the experimental data.  We shift the source of entangled photons away from a quantum GI system's central image plane, and place turbulence near this plane.  This dramatically increases the ghost image contrast.  For turbulence located 482 mm from the crystal, this technique took the observed pattern visibility from  \( \mathcal{V} = 0.14 \pm 0.04\) to  \( \mathcal{V} = 0.29 \pm 0.04\), doubling the system's imaging visibility.

We acknowledge discussions with J.\ H.\ Eberly and support by DARPA DSO InPho grant W911NF-10-1-0404
.


\begin{thebibliography}{99}

\bibitem{PittmanGI}
T.\ B.\ Pittman, Y.\ H.\ Shih, D.\ V.\ Strekalov, A.\ V.\ Sergienko, Phys.\ Rev.\ A, {\bf 52}, R3429 (1995).

\bibitem{ThermalGI}
R.\ S.\ Bennink, S.\ J.\ Bentley, and R.\ W.\ Boyd, Phys.\ Rev.\ Lett., {\bf 89}, 113601 (2002).

\bibitem{Q-C-GI1}
R.\ S.\ Bennink, S.\ J.\ Bentley, R.\ W.\ Boyd, and J.\ C.\ Howell, Phys.\ Rev.\ Lett., {\bf 92}, 033601 (2004).

\bibitem{Q-C-GI2}
A.\ Gatti, E.\ Brambilla, M.\ Bache, and L.\ A.\ Lugiato, Phys.\ Rev.\ Lett., {\bf 93}, 093602 (2004).

\bibitem{Q-C-GI3}
B.\ Erkmen and J.\ H.\ Shapiro, Phys.\ Rev.\ A, {\bf 77}, 043809 (2008).

\bibitem{LenslessImaging}
G.\ Scarcelli, V.\ Berardi, and Y.\ Shih, Appl.\ Phys.\ Lett., {\bf 88}, 061106 (2006).

\bibitem{SuperResolution1}
F.\ Ferri D.\ Magatti, A.\ Gatti, M.\ Bache, E.\ Brambilla, and L.\ A.\ Lugiato, Phys.\ Rev.\ Lett., {\bf 94}, 183602 (2005).

\bibitem{SuperResolution2}
I.\ Vidal, E.\ J.\ S.\ Fonseca, and J.\ M.\ Hickman, Phys.\ Rev.\ A, {\bf 82}, 042827 (2010).

\bibitem{EntDet}
M.\ D'Angelo, Y.\ Kim, S.\ P.\ Kulik, and Y.\ Shih, Phys.\ Rev.\ Lett., {\bf 92}, 233601 (2004).

\bibitem{CompGI}
J.\ H.\ Shapiro, Phys.\ Rev.\ A, {\bf 78}, 061802(R) (2008).

\bibitem{CSGI}
O.\ Katz, Y.\ Bromberg, and Y.\ Silberberg, Appl.\ Phys.\ Lett., {\bf 95}, 131110 (2009).

\bibitem{Turb1}
J.\ Cheng, Opt.\ Exp., {\bf 17}, 7916 (2009).

\bibitem{Turb2}
C.\ Li, T.\ Wang, J.\ Pu, and R.\ Rao, Appl.\ Phys.\ B, {\bf 99}, 599 (2010).

\bibitem{Turb3}
P.\ Zhang, W.\ Gong, X.\ Shen and S.\ Han, Phys.\ Rev.\ A, {\bf 82}, 033817 (2010).

\bibitem{klyshko88}
D.\ N.\ Klyshko, Zh.\ Eksp.\ Teor.\ Fiz.\ {\bf 94}, 82 (1988) [Sov.\ Phys.\ JETP {\bf 67}, 1131 (1988)].

\bibitem{klyshko82}
D.\ N.\ Klyshko, Zh.\ Eksp.\ Teor.\ Fiz.\ {\bf 83}, 1313 (1982) [Sov.\ Phys.\ JETP {\bf 56}, 753 (1982)].

\bibitem{TurbBook}
V.\ I.\ Tatarski, {\it Wave Propagation in a Turbulent Medium}, (McGraw Hill Book Company Inc., New York 1961)

\bibitem{TurbFante}
R.\ L.\ Fante, Prog.\ Opt.\ {\bf 22}, 341-398 (1985).

\bibitem{KolmogorovTurb}
A.\ N.\ Kolmogorov, C.\ R.\ Acad.\ Sci.\ U.S.S.R.\ {\bf 30}, 301-305 (1941).

\end{thebibliography}
\end{document}